# Dynamic characterization of beating sinusoidal wave oscillations in a thin-slice solid-state laser with coupled orthogonally polarized transverse modes

**Kenju Otsuka**

TS$^3$L Research, Yamaguchi 126-7, Tokorozawa 359-1145, Japan
kenju.otsuka@gmail.com

**Abstract**. Dynamic characterization of self-induced beating sinusoidal wave (BSW) oscillations and related phenomena, which were previously explored in a thin-slice dual-polarization solid-state laser operating in the regime of incomplete mode-locking of orthogonally polarized transverse eigenmodes, namely, a quasi-locked state [Laser Physics Letters, 15 (2018) 075001], is performed in terms of the amplitude correlation coefficient and intensity circulation among various orthogonally polarized mode pairs. Polarization-dependent symmetry-breaking featuring a nonreciprocal intensity flow for different orthogonally polarized mode pairs operating in the BSW state are found. The singular state is hidden in a nonlinear system where a perfect mode-locked BSW state with a large amplitude correlation coefficient featuring reciprocal intensity flow is established for a transverse mode pair possessing particular polarization directions that are related to the intensity ratio of orthogonally polarized transverse eigenmodes. Self-organized chaos synchronization is shown to take place for such a singular transverse mode pair subjected to self-mixing modulations.

1. Introduction

Coherent coupling and locking of different transverse modes (namely, transverse mode locking) have been studied in gas lasers [1], semiconductor lasers [2-4], quantum-cascade lasers [5], and solid-state lasers [6, 7]. In addition, laser operation with two orthogonally polarized transverse modes has been studied in the past decade [8, 9]

On the other hand, spontaneous transverse mode locking among different eigenmodes of stable laser resonators, including Hermite-Gauss (HG), Laguerre-Gauss (LG) and Ince-Gauss (IG) transverse modes, has been investigated in thin-slice solid-state lasers (namely, TS$^3$Ls) with coated end mirrors. The mode locking results from actuated saturation-type $\chi^{(3)}$ nonlinearities, i.e., lasing-intensity-dependent refractive index changes, inherent to TS$^3$Ls whose lasing frequency is detuned from the gain peak [10, 11]. Most recently, self-induced beating sinusoidal wave (BSW) oscillations have been observed in quasi-locked states of a thin-slice c-cut Nd:GdVO$_4$ laser with two orthogonally polarized modes, featuring an extremely small mode-splitting and associated modal beat notes [12].

   This paper reports a detailed study of BSW oscillations, paying special attention to the relationship between puzzling small mode-splitting and beat notes as well dynamical characterization of different orthogonally polarized transverse mode pairs in terms of intensity circulation among transverse modes. The observable dynamic quantity of the intensity circulations, which is useful for characterizing complex systems with large degrees of freedom [13, 14], revealed non-reciprocity in the intensity flow

(i.e., symmetry breaking) among interacting orthogonal transverse modes in quasi-locked states. The hidden dynamics in quasi-locked states are elucidated through an intensity circulation analysis, and the effect of small mode-splitting together with associated four-wave mixing is attributed to BSW oscillations in quasi-locked states. Such non-reciprocity is found to disappear for a particular pair of transverse modes, whose polarization directions, $\theta_c$, are determined by the intensity ratio of orthogonally polarized transverse eigenmodes, and synchronized BSW oscillations are shown to feature reciprocal intensity flows among such singular pairs of transverse modes polarized along $\theta_c$. A self-mixing laser Doppler modulation experiment was conducted that confirmed self-organized chaos synchronizations at this singular point.

## 2. Overview of beating sinusoidal-wave oscillations

Before discussing the dynamical characterization of the nonlinear dynamics of a dual-polarization laser, let us review the main point of beating sinusoidal wave (BSW) oscillations in light of new experimental results found by using a thin-slice c-cut Nd:GdVO$_4$ laser operating in orthogonally polarized transverse modes.

*2.1 Stationary characteristics of dual-polarization oscillations.*
The experimental setup is shown in Figure 1(a). A nearly collimated lasing beam from a laser diode (wavelength: 808 nm) was passed through an anamorphic prism pair to transform an elliptical beam into a circular one, and it was focused onto a thin-slice laser crystal by a microscopic objective lens of numerical aperture NA = 0.5. The laser crystal was a 3 mm-diameter clear-aperture, 1 mm-thick, 3 at%-doped c-cut Nd:GdVO$_4$ whose end surfaces were directly coated with dielectric mirrors M$_1$ (transmission at 808 nm > 95%; reflectance at 1064 nm = 99.8%) and M$_2$ (reflectance at 1064 nm = 99%) and whose Fresnel number was $4 \times 10^4$. Lasing patterns were observed with a PbS phototube followed by a TV monitor and an intensity profiler. Lasing optical spectra were measured by a multi-wavelength meter (HP-86120B; wavelength range, 700–1650 nm) for obtaining global views and a scanning Fabry-Perot interferometer (SFPI) (Burleigh SA$^{PLUS}$; 2 GHz free spectral range; 6.6 MHz resolution) for measuring detailed structures. In the case of TS$^3$Ls and monolithic microchip solid-state lasers [11,15,16], the input-output characteristics and the transverse and longitudinal mode oscillation properties depend on the focusing condition of the pump beam on the crystal due to the mode-matching between the pump and lasing mode profiles. In the present experiment, the pump-beam diameter was changed by shifting the laser crystal along the z-axis, as depicted in Figure 1(a).

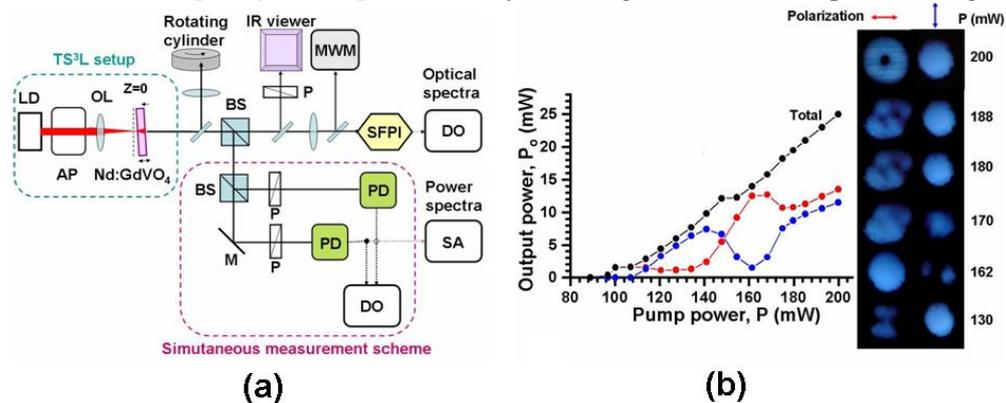

Figure 1. (a) Experimental apparatus. LD: laser diode, AP: anamorphic prism pair, OL: objective lens, BS: beam splitter, P: polarizer, MWM: multi-wavelength meter, SFPI: scanning Fabry-Perot interferometer, PD: photo-diode, DO: digital oscilloscope, SA: spectrum analyser. (b) Input-output characteristics and pump-dependent DPO far-field intensity patterns.

When the laser crystal was adjusted to the pump-beam focus position, i.e. $z = 0$, a $TEM_{00}$ mode oscillation was achieved at the lowest threshold pump power, $P_{th} = 30$ mW, with the highest slope efficiency, $\eta = 50\%$, where the pump and lasing spot sizes were measured to be $w_p = 15$ μm and $w_o = 32$ μm, respectively. Here, the z-dependent pump and near-field lasing spot sizes were measured with the aid of a laser beam profiler (Coherent, Beam Master 3Si). $w_p$ increased as the laser crystal was shifted away from the pump-beam focus along the z-axis (i.e., $z > 0$) and $P_{th}$ was increased. When $w_p$ exceeded about 80 μm, dual-polarization oscillations (DPOs) were observed, starting from $TEM_{00}$ at the threshold and leading to various transverse modes. When the pump position was precisely changed by moving the laser sample along the x-axis or y-axis with an accuracy of 10 μm and a small tilt of the 1-mm-thick cavity of $|a| \leq 1.5°$ was made with an accuracy of $0.3°$ as depicted in the inset, several transverse modes appeared at fixed $w_p$. These modes within the pumped region reflected the roughness of the polished surface [11] and the effect of 'tilted' wide-aperture pumping on the radial saturated inversion population distribution along the thickness direction of the 1-mm-thick crystal, resulting in oblique gain guiding against the lasing axis [17].

The input-output characteristics and the corresponding DPO modal structural changes are shown in Figure 1(b) versus the pump power. Here, $w_p = 80$ μm, and the first lasing mode, i.e. the preceding mode, starts from $TEM_{00}$ at a threshold pump power of $P_{th} = 92$ mW, leading to an Ince-Gauss mode, $IG^e_{2,2}$, while second lasing mode $HG_{0,0}$ appears with increasing pump power, and simultaneous oscillation of these two orthogonally polarized modes takes place at $P \geq 160$ mW. Upon further increase of the pump power, the $IG^e_{2,2}$ mode makes a structural change to a 'doughnut' mode, while the $HG_{0,0}$ mode remains the orthogonally polarized partner. Here, the transverse spatial hole-burning effect was such that the second lasing mode had less transverse cross-saturation of the population inversions against the preceding mode [11]. The DPO shown in Figure 1 (b) was reproducibly obtained by adjusting the crystal and pump positions [x, y, z] as well as the tilt angle [11]. The 'doughnut' mode has been identified as the hybrid $TEM_{0,1}^*$, as reported in [12].

The output DPO lasing beam was passed through a polarizer, and the polarization-dependent changes in the far-field patterns were observed. Experimental results corresponding to $IG^e_{2,2} - HG_{0,0}$ and $TEM_{0,1}^* - HG_{0,0}$ pairs are shown in Figures 2(a) and 2(b), together with the total output intensity profile. The global oscillation spectrum measured by the multi-wavelength meter indicated a single longitudinal mode at a wavelength of $\lambda = 1065.58$ nm, which corresponds to the $\sigma_1$-polarization transition line, $^4F_{3/2}(1) \rightarrow {}^4I_{11/2}(1)$ [18], while the adjacent longitudinal mode separated by $\Delta\lambda = \lambda^2/2nL = 0.258$ nm (n = 2.1981: refractive index, L = 1 mm: cavity length) was not observed.

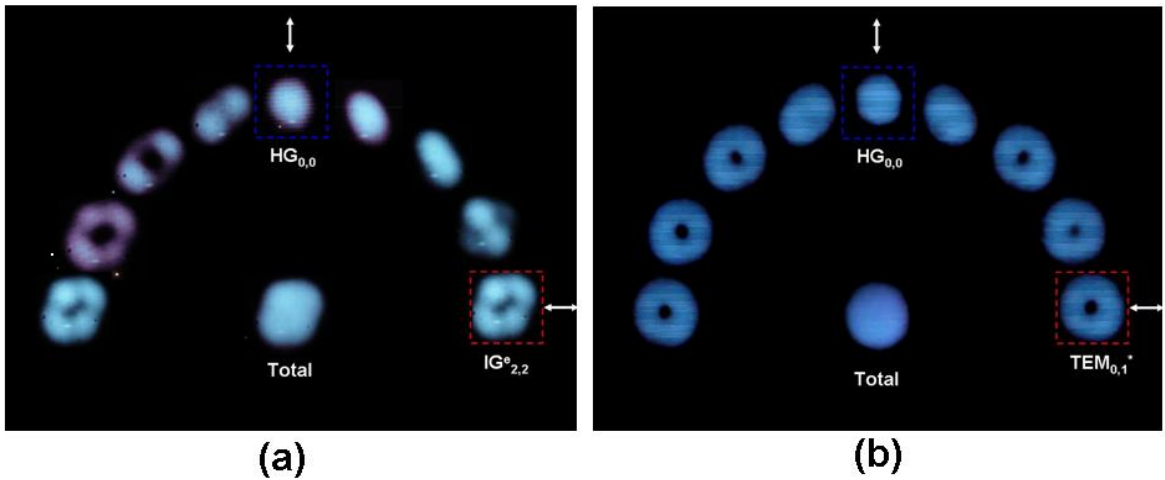

Figure 2. Polarization dependent structural changes of far-field patters. Pump power, P: (a) 175 mW, (b) P = 200 mW. See video clips for polarization-dependent pattern changes.

*2.2 Beating sinusoidal-wave oscillations and their optical and power spectra*

Figures 3(a)-(b) show an example of the temporal evolution of modal intensities of orthogonally polarized $IG^e_{2,2}$ and $HG_{0,0}$ modes and the corresponding power spectra around the relaxation oscillation frequency regime, $f_1 = (1/2\pi)[(P/P_{th} - 1)/\tau\tau_p]^{1/2}$ ($\tau$: fluorescence lifetime; $\tau_p$: photon lifetime), and the lower frequency relaxation noise peak, $f_2$, which reflects the cross-saturation dynamics of the population inversions of the interacting orthogonally polarized transverse modes. Here, two InGaAs photodetectors (New Focus 1611; 3-dB bandwidth: 30 kHz-1 GHz) and a digital oscilloscope (Tektronics TDS 3052; bandwidth: 500 MHz) were used. The AC-coupled output voltages from the photodetectors shown in Figure 3(a) are proportional to the lasing output intensity fluctuations. The observation time in the whole experiment described below was 100 μs for $10^5$ data.

The detailed optical spectra corresponding to the orthogonally polarized $IG^e_{2,2}$-$HG_{0,0}$ mode operations in Figure 2(a), as measured by the SFPI, are shown in Figure 3(c). Apart from the complete transverse mode locking of the orthogonally polarized modes, peculiar, strongly asymmetric optical mode spectra, each consisting of dominant peaks, $E_1$, $E_2$, at optical frequencies, $\nu_1$, $\nu_2$, and weak subsidiary peaks, $E_3$, $E_4$, at $\nu_3$, $\nu_4$ appear for the orthogonally polarized $IG^e_{2,2}$ and $HG_{0,0}$ modes, as shown in of Figure 3(c). Here, oscillation frequency detuning on the order of $\Delta\nu = \nu_1 - \nu_2 \cong 400$ MHz, which is much smaller than the difference between the 'cold-cavity' resonance frequencies of the $IG^e_{2,2}$ and $HG_{0,0}$ modes (on the order of several GHz) estimated from their combined mode numbers [11], remains between the orthogonally polarized transverse eigenmodes of the $TS^3L$ cavity. In short, lasing frequencies, $\nu_1$ and $\nu_2$, are shifted from cold-cavity resonance frequencies of $IG^e_{2,2}$ and $HG_{0,0}$ modes, respectively, due to the effect of transverse-mode locking. On the other hand, the small peak $E_3$ is excited at $\nu_3$ ($\cong \nu_2$) in the subsidiary region of the $IG^e_{2,2}$ modal field, as indicated by the arrow. Accordingly, the other weak field $E_4$ is considered to be excited at $\nu_4$ through the four-wave mixing process due to the inherent $\chi^{(3)}$ laser nonlinearity satisfying $\nu_4 = \nu_1 + (\nu_2 - \nu_3)$ with the phase-matching condition of $k_4 = k_1 + k_2 - k_3$, where k is the corresponding momentum of $\nu_I$ [12]. Because of the anomalous dispersion effect around $\nu_1$, which is detuned from the gain peak, the scattered field $E_4$ is considered to be pushed slightly out toward the higher frequency side in the case of Figure 3(c) [19, 20]. As a result, the frequency difference between the peaks, $\Delta\nu_A = \nu_1 - \nu_3$ and $\Delta\nu_B = \nu_4 - \nu_2$, become slightly different by $f_b = \Delta\nu_B - \Delta\nu_A$ for orthogonally polarized fields. In other words, the observed orthogonally polarized transverse modes are not independent, and they form partially coherent fields, namely, a quasi-locked state, through mutual nonlinear interaction of closely spaced $IG^e_{2,2}$ and $HG_{0,0}$ modes [5,12].

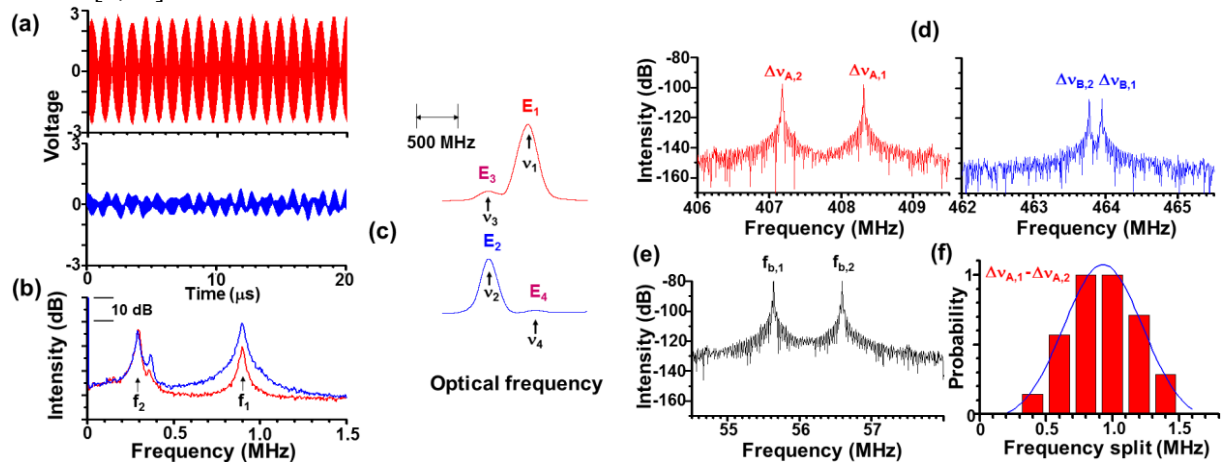

Figure 3. Brief summary of BSW oscillations in quasi-locked states. (a) Polarization-resolved waveforms for orthogonally polarized eigenmodes. (b) Polarization-resolved power spectra around relaxation oscillation frequency. (c) Polarization-resolved optical spectra for eigenmodes. (d) High-frequency modal beat notes. (e) Resultant BSW frequencies. (f) Histogram of mode-splitting of $\Delta\nu_A$.

Finally in this overview section, let us examine typical examples of the power spectra of DPO in a quasi-locked state for different frequency regimes responsible for BSW oscillations. The frequency splittings around $\Delta\nu_A$ and $\Delta\nu_B$ in Figure 3(d) correspond to the small mode splittings of the fields, $E_3$ and $E_4$, respectively. The resulting polarization-resolved power spectra in the beating frequency regime, $f_{b,1}$ and $f_{b,2}$, are shown in Figure 3(e). Here, a generic relation for the beating frequencies $f_{b,1}$ and $f_{b,2}$: $\Delta f_b = |\Delta f_A - \Delta f_B| = \pm(f_{b,2} - f_{b,1})$; $f_{b,2} > f_{b,1}$ for $\Delta f_A > \Delta f_B$ and vice versa, is established for the observed quasi-locked states, where $f_{b,1} = \Delta\nu_{B,1} - \Delta\nu_{A,1}$ and $f_{b,2} = \Delta\nu_{B,2} - \Delta\nu_{A,2}$ [12]. The beating frequency $\Delta f_b$ determined by the mode-splitting of $\Delta f_A$ and $\Delta f_B$ fluctuates over time even at a fixed pump power. A histogram of the fundamental $E_3$-mode splitting frequencies, $\Delta\nu_A$, is shown in Figure 3(f).

The physical mechanism behind the extremely small splitting of $\Delta\nu_A$ and the associated $\Delta\nu_B$-splitting less than 1 MHz is still a puzzling subject despite that the BSW oscillations could be interpreted in terms of the combined effect of such a mode-splitting and the associated four-wave mixing. Further discussion on this subject is presented in *3.3*.

Essentially the same phenomena were observed for the orthogonally polarized TEM$_{0,1}^*$ and HG$_{0,0}$ modes when the pump power was increased.

## 3. Dynamical characterization of polarization-dependent modal interactions

This section shows the polarization-dependent power spectra and modal interactions among different pairs of orthogonally polarized fields, which were examined for orthogonally polarized pairs of IG$^e_{2,2}$ - HG$_{0,0}$ modes in Figure 2(b).

*3.1 Measurement of polarization-dependent power spectra of BSW oscillations and intensity probability distributions*

Polarization-dependent dynamics, which give a foundation for investigating nonlinear interactions between various pairs of orthogonally polarized transverse modes, were measured for the orthogonally polarized IG$^e_{2,2}$-HG$_{0,0}$ mode operations in Figure 2(a).

Figures 4(a) shows the polarization-dependent power spectral intensity, $I_{RO}(\theta)$, at the intrinsic relaxation oscillation frequency, $f_1$, and the lower frequency relaxation noise peak, $f_2$, in Figure 3(b), while Figure 4(b) shows the polarization-dependent power spectral intensity, $I_b(\theta)$, at the beating frequencies, $f_{b,1}$ and $f_{b,2}$, which are relevant to the BSW oscillations. Note that $f_2$ relaxation oscillations resulting from the inherent transverse cross-saturation of population inversions among modes [21] are strongly suppressed at the critical polarization directions, $\theta_c = 50.5^o$ and $131.5^o$, as shown in Figure 4(c). Here, the critical angles are given by $\theta_c \cong \arctan(\pm 1/r)$ when subsidiary peaks are much smaller compared with the dominant peaks, i.e., $E_3 \ll E_1$ and $E_4 \ll E_2$, where $r = (I_2/I_1)^{1/2}$ is the field amplitude ratio of DPO eigenmodes in Figure 1(b), where the field components of the orthogonally polarized eigenmodes along $\theta_c$ coincide and the effective modal gain for one mode coincides with that of the other mode [12]. The critical polarization state, at which $f_2$-noise vanishes and such a pair of transverse modes behaves just like a single mode laser through coherent field coupling among modes, is referred to as the "vanishing polarization state" hereafter. In the absence of coherent modal field coupling, on the other hand, antiphase dynamics arise featuring $f_2$ noise even if a pair of modes possesses the equal gain [22, 23]. Figure 4(d) shows histograms of intensity fluctuations for different polarization directions, $\theta$. A peculiar non-Gaussian 'witches hat' type of distribution that corresponds to the BSW oscillations of the IG$^e_{2,2}$ mode at $\theta = 0^o$ gradually changes into the Gaussian intensity distribution of the HG$_{0,0}$ mode as the polarizer angle approaches $\theta = 90^o$ [12].

The generic properties of this state will be discussed in the following sections in connection to the intensity circulation.

*3.2 Intensity circulation analysis of orthogonally polarized modes*

To identify the nonlinear interaction among different orthogonally polarized mode pairs, let us introduce the dynamic quantity of intensity circulation and examine their dynamic independence.

Here, the observable dynamic quantity of intensity circulation is defined as [13, 14]:

$$I_{i,j} = I_i \dot{I}_j - \dot{I}_i I_j. \tag{1}$$

$I_{i,j} > 0$ (< 0) implies an intensity flow from mode *i* (*j*) to mode *j* (*i*).

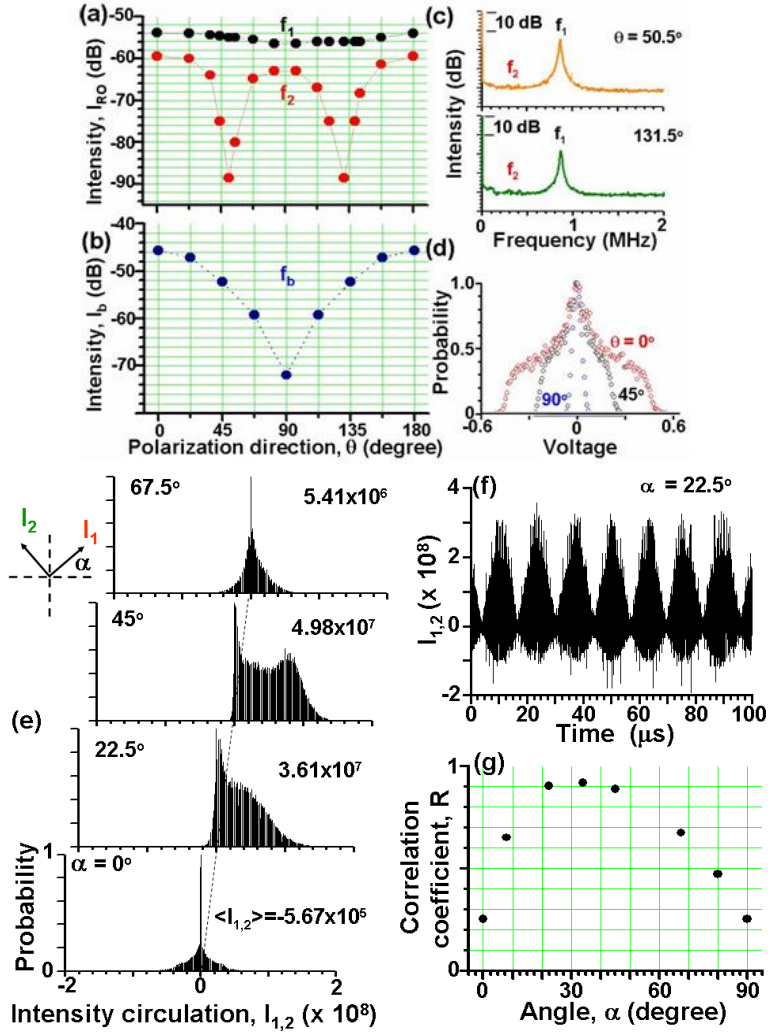

Figure 4. Polarization-dependent peak power spectral intensities in regions of (a) relaxation oscillation frequencies, $f_1$ and $f_2$, and (b) BSW frequency, $f_b$. (c) Power spectra around $f_1$ and $f_2$. (d) Probability distribution normalized by the peak value for polarization directions, $\theta = 0°$, $45°$, and $90°$. (e) Probability distributions of intensity circulations for different pairs of orthogonally polarized modes. (f) Intensity circulation waveforms for $\alpha = 22.5°$. (g) Amplitude correlation coefficient as a function of rotation angle, $\alpha$. Pump power, $P = 175$ mW.

The experimental scheme for accurate simultaneous measurement of two interacting orthogonally polarized modes is indicated by the dashed line in the lower part of Figure 1(a). To ensure accuracy, the total output beam in the BSW oscillations was divided into two beams having equal intensity and focused onto individual InGaAs photo-detectors with the polarizers being removed. Here, the distances to the two photo-detectors and the lengths of cable to the input channels of the digital oscilloscope were adjusted to be the same; in addition, the delay time difference between the two input channels was kept within the sampling period of 1 ns on the basis of the phase difference between the two BSW waveforms. This is because the intensity circulation critically depends on the relative phase difference of the BSW oscillations for the interacting orthogonal transverse modes.

Each beam was then passed through a polarizer and intensity circulation analysis was carried out on several pairs of orthogonally polarized transverse modes. The normalized probability distributions of the intensity circulation $I_{1,2}(t)$ for different pairs are shown in Figures 4(e), where nonreciprocal intensity flows representing the symmetry breaking are formed in quasi-locked states except for the orthogonally polarized $IG^e_{2,2}$ and $HG_{0,0}$ eigenmode pair at $\alpha = 0°$ which exhibits reciprocal intensity flow. An example of $I_{1,2}(t)$ for $\alpha = 22.5°$ is shown in Figure 5(f), where periodic bursting occurs at the beating frequency, $\Delta f_b$. The average intensity circulation, $<I_{1,2}>$, in the figure increases with increasing $\alpha$ toward $\alpha \cong 45°$. On the other hand, the relative phase difference among a pair of BSWs for orthogonally polarized modes was verified to arise and increase with increasing $\alpha$ toward $\alpha \cong 45°$. Therefore, the non-reciprocity can be attributed to such a nonlinear 'built-in' phase difference reflecting the failure of the complete transverse mode locking.

The amplitude correlation coefficient, $R = \Sigma_i (I_{1,i} - <I_1>)(I_{2,i} - <I_2>) / [\Sigma_i (I_{1,i} - <I_1>)]^{1/2}[\Sigma_i (I_{1,i} - <I_1>)]^{1/2}$ is plotted as a function of the rotation angle, $\alpha$, in Figure 5(g). The amplitude correlation coefficient, $R(\alpha)$, is low for the orthogonally polarized $IG^e_{2,2}$ and $HG_{0,0}$ eigenmode pair at $\alpha = 0°$, while it peaks around $\alpha \cong 45°$. This is reasonable because individual output intensities of the pair of orthogonally polarized modes approach similar BSW waveforms with a 'witches hat' type of probability distribution (Figure 4(d)) as $\alpha$ approaches $\alpha \cong 45°$. In the case of complete transverse mode-locking of odd- and even IG modes, on the contrary, the reciprocal intensity flow between noise-driven relaxation oscillations with $R \cong 1$, is established for any pair of interacting transverse modes with different polarizations [10].

Essentially the same nonlinear dynamics were observed for a pair of orthogonally polarized $TEM_{0,1}$ and $HG_{0,0}$ modes when the pump power $P \geq 190$ mW (Figure 2(b)).

*3.3 Physical significance of intensity circulations*
The intensity circulation analysis demonstrated so far is useful for dynamic characterization of coupled degrees of freedom in complex systems, which are generally similar to gain circulations in multimode lasers [13]. Furthermore, power spectral analyses of the intensity circulations can reveal hidden dynamics embedded in the complex system.

Here, a power spectral analysis is performed, paying special attention to the mechanism of BSW oscillations: mode-splitting mediated four-wave mixing and modal beating in quasi-locked states. The results, which correspond to the BSW oscillations summarized in Figure 3, are shown in Figure 5. In the low-frequency domain around the relaxation oscillation frequency, a strong peak corresponding to the beat frequency, $\Delta f_b$, which cannot been recognized in the power spectra of the BSW oscillations themselves, is manifested in Figure 5(a). In the high-frequency domain around $\Delta\nu_A$ and $\Delta\nu_B$, peculiar $\Delta f_b$ sidebands appear besides the split peaks separated by $\Delta f_A = \Delta\nu_{A,1} - \Delta\nu_{A,2}$ and $\Delta f_B = \Delta\nu_{B,1} - \Delta\nu_{B,2}$, as indicated by the arrows in Figure 5(b).

Let us discuss the physical origin of these peculiar $\Delta f_b$ sidebands. The beating frequencies are given by $f_{b,1} = \Delta\nu_{B,1} - \Delta\nu_{A,1}$, $f_{b,2} = \Delta\nu_{B,2} - \Delta\nu_{A,2}$, and the resultant BSW oscillations occur at $\Delta f_b = |\Delta f_A - \Delta f_B|$, as described in section *2.2*. This strongly implies that the split fields, $E_{3,1}$ ($E_{3,2}$) and $E_{4,1}$ ($E_{4,2}$), are coupled coherently to produce a beat note at $f_{b,1}$ ($f_{b,2}$), assuming a four-wave mixing process wherein $E_{4,1}$ is created from $E_1$, $E_2$, and $E_{3,1}$, while $E_{4,2}$ is created from $E_1$, $E_2$, and $E_{3,2}$ (Figure 5(c)). Similarly, the peculiar $\Delta f_b$ sidebands are such that they appear inside or outside of the beat signals separated $\Delta\nu_{A,2}$ and $\Delta\nu_{B,2}$, as indicated by the arrows in Figure 5(d) in accordance with the four-wave mixing process among the split fields and their $\Delta f_b$-sideband fields.

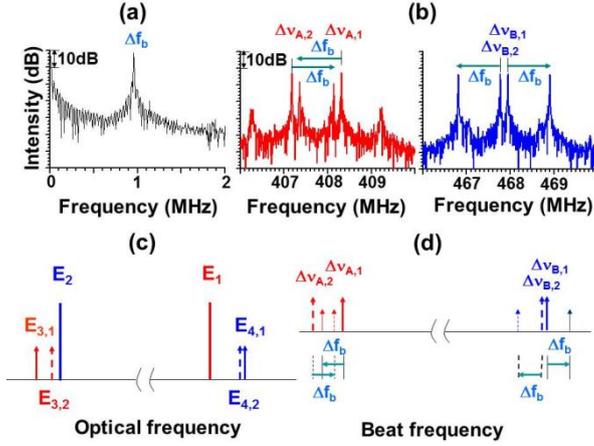

Figure 5. Power spectral analyses of intensity circulations in the frequency regions of (a) beating frequency, $\Delta f_b$, and modal beat notes, $\Delta \nu_A$ and $\Delta \nu_B$. Schematic illustrations of (c) lasing mode spectrum and (d) modal beat notes.

## 4. Vanishing antiphase dynamics

In this section, nonlinear dynamics of a pair of transverse modes polarized along the directions, $\theta_c$, are examined, featuring the intensity circulation and amplitude correlation between these particular transverse modes, in BSW as well as chaotic oscillations.

*4.1. Synchronized BSW oscillations*
Simultaneous measurements of two output signals polarized along $\theta_c$ were carried out using the same scheme as described in section *2.2*. The results for $IG^e_{2,2}$-$HG_{0,0}$ and $TEM_{0,1}^*$-$HG_{0,0}$ pairs in Figures 2(a)-(b) are summarized in Figures 6(a) and 6(b), which show magnified views of the waveforms and amplitude correlation plots. It is apparent that the two BSW waveforms are completely synchronized; the amplitude correlation coefficient reaches $R = 0.995$ and $0.997$, respectively. The corresponding long-term temporal evolutions of the intensity circulations and their probability distributions for these cases are shown in in Figures 6(c) and 6(d).

It is apparent that the reciprocal intensity flows are recovered for vanishing polarization states; i.e., symmetry breaking recovers for orthogonally polarized mode pairs shown in Figures 4(e). In short, the two modes polarized along $\pm\theta_c$ behave like an "all-in-one" coherent mode whose output power spectrum exhibits a single relaxation-oscillation frequency peak at $f_1$.

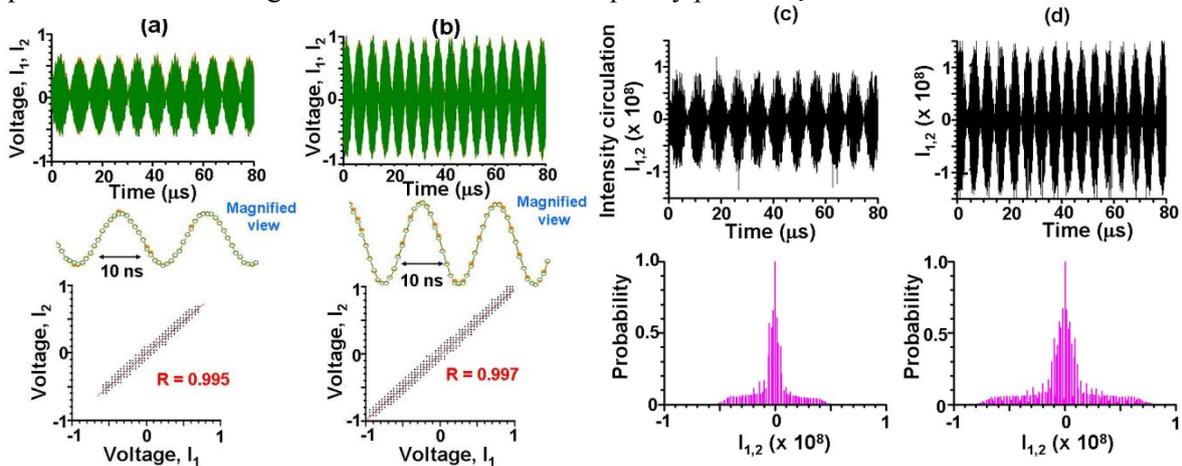

Figure 6. Synchronized BSW oscillations for a pair of transverse modes in vanishing polarization states. (a), (c): P = 175 mW, $\theta_c = \pm 50.5^o$. (c), (d): P = 200 mW, $\theta_c = \pm 47.5^o$.

*4.2 Self-organized chaos synchronization with self-mixing modulations*

Synchronization phenomena of chaotic oscillators [24] are encountered in physical, chemical, and biological systems, and in laser systems. In laser physics, a variety of chaos synchronizations have been reported in coherently coupled solid-state lasers [25, 26] and semiconductor lasers [27, 28]. Here, a self-mixing modulation experiment [26] employing an extremely sensitive laser Doppler feedback scheme with thin-slice solid-state lasers having a large fluorescence-to-photon lifetime ratio [29-31] was carried out to investigate chaos synchronizations in the quasi-locked states involving orthogonally polarized transverse modes, paying special attention to the generic nature of the vanishing polarization states against external perturbations.

Part (50%) of the output beam was focused onto a rotating Al-cylinder with rough surfaces, as depicted in Figure 1(a), and the rest of the beam was divided into two beams for simultaneous measurement of the two signals in vanishing polarization states. When the Doppler-shift frequency, $f_D = 2v/\lambda$ (v: moving speed along the laser axis), was tuned around the relaxation oscillation frequency, i.e., $f_D \sim f_1$, chaotic relaxation oscillations were easily brought about, while BSW oscillations at $\Delta f_B$ occurring without modulations were suppressed. In short, the power spectral analysis, where sinusoidal oscillations at the beating frequency at $f_b = \Delta\nu_B - \Delta\nu_A$ superimposed on the chaotic waveforms were very rarely observed, indicates that the mode-splitting of $\Delta\nu_A$ and $\Delta\nu_B$, as in Figure 3(c), was strongly disturbed by the large-amplitude chaotic fluctuations in intensity.

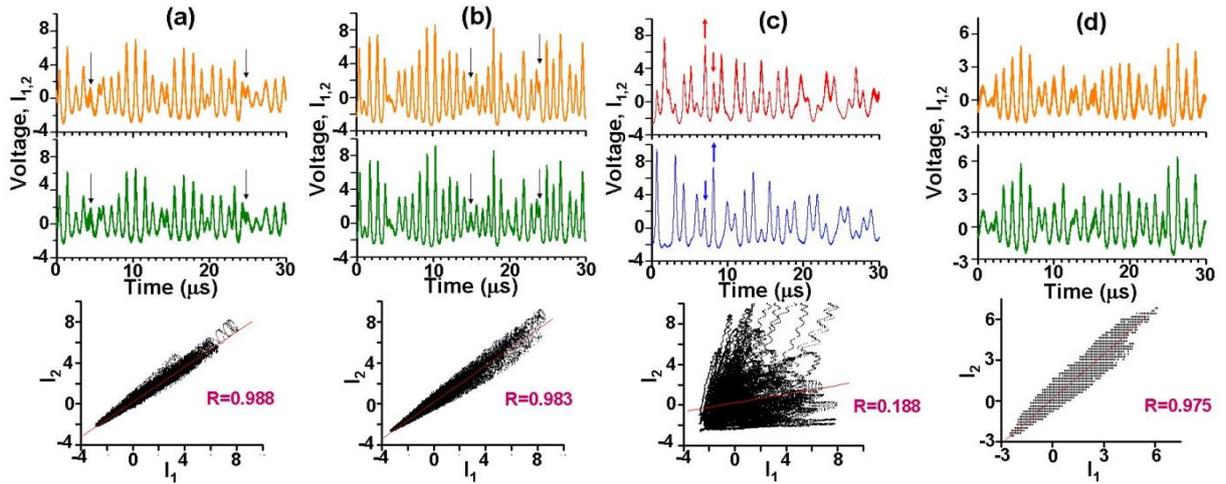

Figure 7. (a), (b) Synchronized chaotic oscillations for a pair of transverse modes in vanishing polarization states; the pump power was 175 mW in (a) and 200 mW in (b). (c) Antiphase chaotic oscillations for the $IG^e_{2,2}$ and $HG_{0,0}$ eigenmode pair. P = 175 mW. (d) Chaos synchronization when one mode in the vanishing polarization state was subjected to self-mixing modulation. P = 175 mW.

Even in the chaotic regime, the generic nature of the vanishing polarization states is preserved such that the pair of vanishing polarization states behaves in all-in-one coherent mode, exhibiting self-organized chaos synchronization. Typical examples for the $IG^e_{2,2}$-$HG_{0,0}$ and $TEM_{0,1}^*$-$HG_{0,0}$ DPO cases are shown in Figures 7(a) and (b), where substantially high amplitude correlation coefficients of R = 0.988 and 0.983 are attained even in such chaotic regimes only for the pair of transverse modes polarized along $\theta_c$ directions, where sinusoidal waveforms at $f_b = \Delta\nu_B - \Delta\nu_A$ occasionally appear near the averaged output intensity, i.e., near zero AC voltage in Figures 7(a) and (b) (indicated by the arrows). When $\theta$ was shifted slightly from $\theta_c$, on the other hand, chaos synchronization was readily suppressed. If we look at the modal outputs for a pair of transverse eigenmodes, i.e., $IG^e_{2,2}$ and $HG_{0,0}$, in the chaotic regime, chaos synchronizations failed and inherent antiphase pulsations took place instead, as indicated by the arrows in Fig. 7(c), where R = 0.188.

Finally, it should be noted that the high degree of chaos synchronization was established between transverse modes polarized along $\pm\theta_c$ directions when one mode was subjected to self-mixing modulation, where part of the polarized output along $+\theta_c$ (or $-\theta_c$) in the simultaneous measurement scheme in Figure 1(a) was focused on the rotating cylinder. An example is shown in Figure 7(d). This sender-receiver type of synchronization shows potential for a secure metrology system, where a self-mixing signal created by the output beam toward a target [30], which is polarized along $\pm\theta_c$, can be accurately retrieved solely by the output beam polarized along $\mp\theta_c$.

## 5. Summary

Dynamical characterization of quasi-locked states among orthogonally polarized transverse modes in a thin-slice c-cut isotropic Nd:GdVO$_4$ polarization vector laser were carried out, focusing on nonlinear dynamics of "vanishing polarization states" formed from a pair of transverse modes polarized along well-defined directions determined by the intensity ratio of orthogonally polarized eigenmodes of the cavity. The pair of transverse modes in the vanishing polarization states, which act as a single-mode laser free from the transverse cross-saturation of population inversions among eigenmodes, behaves as if it is in all-in-one coherent mode. The vanishing polarization states were found to show synchronized beating sinusoidal oscillations as well as self-organized chaos synchronizations.

## References

[1]  Brambilla M, Battipede F, Lugiato L A, Penna V, Prati F, Tamm C and Weiss C O (1991): Transverse laser patterns. I. Phase singularity crystals *Phys. Rev. A* **43** 5090-5113
[2]  Tan G L, Mand R S and Xu J M (1997): Self-consistent modeling and beam instabilities in 980-nm fiber pump lasers *IEEE J. Quantum Electron.* **33** 1384-1395
[3]  Fu X, Tan G L, Gordon R and Xu J M (1998): Third-order nonlinearity induced lateral-mode frequency locking and beam instability in the high power operation in narrow-ridge semiconductor lasers *IEEE J. Quantum Electron.* **8** 1447-1454
[4]  Scheuer J and Orenstein M (1999): Optical vortices crystals: spontaneous generation in nonlinear semiconductor microcavities *Science* **285** 230-233
[5]  Yu N, Diehl L, Cubukcu E, Bour D, Corzine S, Höfler G, Wojcik A K, Crozier K B, Belyanin A and Capasso F (2009): Coherent coupling of multiple transverse modes in quantum cascade lasers *Phys. Rev. Lett.* **102** 013901
[6]  Staliunas K, Tarroja M F H and Weiss C O (1993): Transverse mode locking, anti-locking and self-induced dynamics of class-B lasers *Opt. Commun.* **102** 69-75
[7]  Otsuka K and Chu S C (2009): Generation of vortex array beams from a thin-slice solid-state laser with shaped wide-aperture laser-diode pumping *Opt. Lett.* **34** 10–12
[8]  Oron R, Blit S, Davidson N and Friesem A (2000): The formation of laser beams with pure azimuthal or radial polarization *Appl. Phys. Lett.* **77** 3322-334.
[9]  Oron R, Shimshi L, Blit S, Davidson N, Friesem A and Hasman E (2002); Laser operation with two orthogonally polarized transverse modes *Appl. Opt.* **41** 3634-3637
[10] Otsuka K, Chu S C, Lin C C, Tokunaga K and Ohtomo T (2009) Spatial and polarization entanglement of lasing patterns and related dynamic behaviors in laser diode-pumped solid-state lasers *Opt. Express* **17** 21615-21627
[11] Otsuka K and Chu S C (2017): Spontaneous generation of vortex and coherent vector beams from a thin-slice c-cut Nd:GdVO$_4$ laser with wide-aperture laser-diode end pumping: Application to highly sensitive rotational and translational Doppler velocimetry *Laser Phys. Lett.* **14** 075002

[12] Otsuka K (2018): Self-induced beating sinusoidal wave oscillations and polarization-dependent nonlinear dynamics in a thin-slice solid-state laser with coupled orthogonally polarized transverse modes *Laser Phys. Lett.* **15** 075001

[13] Otsuka K and Aizawa Y (1994): Gain circulation in multimode lasers *Phys. Rev. Lett.* **72** 2701-2704

[14] Otsuka K, Sato Y and Chern J L (1996): Grouping of antiphase oscillations in modulated multimode lasers *Phys. Rev. A* **54** 4464-4472

[15] Naidoo D, Harfouche A, Fromager M, Ait-Ameur K and Forbes A (2016): Emission of a propagation invariant flat-top beam from a microchip laser *J. Luminescence* **170** 750-754

[16] Naidoo D, Fromager M, Ait-Ameur K and Forbes A (2015): Radially polarized cylindrical vector beams from a monolithic microchip laser *Opt. Engineering* **54** 111304

[17] Dong J, Bai S C, Liu S H, Ueda K and Kaminskii A A (2016): A high repetition rate passively Q-switched microchip laser for controllable transverse laser modes *J. Opt.* **18** 055205

[18] Anderson F G, Summers P L, Weidner H, Hong P and Peal E E (1994): Interpretive crystal-field parameters: application to $Nd^{3+}$ in $GdVO_4$ and $YVO_4$, *Phys. Rev. B* **50** 14802

[19] Kawaguchi H, Inoue K, Matsuoka T and Otsuka K (1985): Bistable output characteristics in semiconductor laser injection locking *IEEE J. Quantum Electron.* **21** 1314–1317

[20] Chern J L, Kubota T, Lim T S and Otsuka K (2002): Stokes like emissions from laser-diode-pumped microchip neodymium-doped solid-state lasers *J. Opt. Soc. Am. B* **19** 1668-1675

[21] Ko J Y, Lin C C, Otsuka K, Miyasaka Y, Kamikariya Ko, Nemoto K, Ho M C and Jiang I M (2007): Experimental observations of dual-polarization oscillations in laser-diode-pumped wide-aperture thin-slice $Nd:GdVO_4$ lasers *Opt. Express* **15** 945-954

[22] Mandel P, Otsuka K and Pieroux D (1993): Transient and modulation dynamics of a multimde Fabry-Perot laser *Opt. Commun.* **100** 341-350

[23] Otsuka K (2000): *Nonlinear dyanmics in optical complex systems* (Springer Netherland)

[24] Pecora L M and Carroll T L (1990): Synchronization in chaotic systems *Phys. Rev. Lett.* **64** 821-824

[25] Roy R and Thornburg, Jr. K S (1994): Experimental synchronization of chaotic lasers *Phys. Rev. Lett,* **72** 2009-2012

[26] Otsuka K, Kawai R, Hwong S L, Ko J Y and Chern J L (2000): Synchronization of mutually coupled self-mixing modulated lasers *Phys. Rev. Lett.* **84** 3049-3052

[27] Heil T, Fischer I, Elsässer W, Mulet J and Mirasso C R (2001): Chaos synchronization and spontaneous symmetry-breaking in symmetrically delay-coupled semiconductor lasers *Phys. Rev. Lett.* **86** 795-798

[28] Liu Y, Takiguchi Y, Davis P and Aida T (2002): Experimental observation of complete chaos synchronization in semiconductor lasers *Appl. Phys. Lett.* **80** 4306

[29] Otsuka K (1979): Effects of external perturbations on $LiNdP_4O_{12}$ lasers *IEEE J. Quantum Electron.* **QE-15** 655-663

[30] Otsuka K (2011): Self-mixing thin-slice solid-state laser metrology *Sensors* **11** 2195-2245

[31] Zhu K, Chen H, Zhang S, Shi Z, Wang Y and Tan Y (2019): Frequency-shifted optical feedback technologies using a solid-state microchip laser *Appl. Sci.* **9** 109-1~27